\documentclass[twocolumn,showpacs,preprintnumbers,amsmath,amssymb,showkeys]{revtex4}
\usepackage[dvips]{graphicx}% Include figure files
\usepackage{dcolumn}% Align table columns on decimal point
\usepackage{bm}% bold math

\begin{document}

%\preprint{APS/123-QED}

\title{Chimeras in networks of planar oscillators}% Force line breaks with \\

\author{Carlo R. Laing}
\email{c.r.laing@massey.ac.nz}
\affiliation{ IIMS, Massey University, Private Bag 102-904 NSMC, Auckland, New Zealand}

\date{\today}

\begin{abstract}
Chimera states occur in networks of coupled oscillators, and are characterized by having some fraction
of the oscillators perfectly synchronized, while the remainder are desynchronized. Most chimera states
have been observed in networks of phase oscillators with coupling via a sinusoidal function of phase differences,
and it is only for such networks that any analysis has been performed. Here we present the first analysis of chimera
states in a network of planar oscillators, each of which is described by both an amplitude and a phase.
We find that as the attractivity of the underlying periodic orbit is reduced chimeras are destroyed
in saddle-node bifurcations, and supercritical Hopf and homoclinic bifurcations of chimeras also occur. 
\end{abstract}

    % PACS, the Physics and Astronomy Classification Scheme.
                             
 % Use showkeys class option if keyword display desired

\maketitle

Networks of coupled oscillators and their synchronization properties
have been studied for many years~\cite{pikovskybook03,str03}. 
One particular class of interest involves
phase oscillators, where each oscillator is described by a single angular variable~\cite{acebon05,str00}. 
The use of such phase models is justified when the attraction to an underlying limit cycle is ``strong''
relative to the effects of other oscillators in the network~\cite{pikovskybook03,acebon05,kuramotobook84}.
Recently a number of investigators have studied ``chimera'' states in networks of phase 
oscillators~\cite{pikros08,kurbat02,lai09B,lai09A,abrmir08,abrstr04,abrstr06,setsen08,omemai08,shikur04,kawa07,kurshi06,marlai10},
in which some fraction of the oscillators synchronize while the remainder run freely, even though the
oscillators may be indentical. Early analyses of these 
states~\cite{kurbat02,abrstr04,abrstr06,setsen08,omemai08,shikur04,kurshi06,kawa07,marlai10}
used a self-consistency
argument
which can be traced back to~\citet{kuramotobook84} to show existence of chimeras. Later 
work~\cite{lai09A,lai09B,abrmir08,mar10}
used the remarkable ansatz of Ott and
Antonsen~\cite{ottant08,ottant09} to derive differential equations governing the evolution of order parameters of the
systems under study, allowing one to determine the stability of chimera states and the bifurcations 
they may undergo.

It has long been known that networks of identical phase oscillators, coupled through a sinusoidal function
of phase differences, have non-generic behaviour~\cite{pikros08,watstr93,watstr94,marmir09}. Most chimera
states have been observed in such idealized networks, and in order to determine whether chimeras might be observed
in real physical systems one should investigate their robustness with respect to, for example, heterogeneity
in intrinsic frequencies, or variations in oscillator amplitude. The first issue has already been 
addressed~\cite{lai09A,lai09B}, and here we investigate the second.

Several authors have observed chimeras in networks of oscillators described 
by more than one variable~\cite{sak06,kawa07,kurshi06,kurbat02,shikur04,kurshi03}, so they are known to exist,
but these authors have either provided no analysis, or have reduced their (identical) oscillators to phase oscillators in
order to analyse their dynamics using the approaches mentioned above. 
In this Letter we give the first analysis of a chimera state in a network of
planar oscillators in which the reduction to phase oscillators is not performed. 
%By allowing not only the phase
%but the amplitude of each oscillator to vary we introduce the possibility that this variation in amplitude
%could cause bifurcations of chimeras.

The model we consider is 
\begin{eqnarray}
   \frac{dX_j}{dt} & = & i\omega X_j+\epsilon^{-1}\left\{1-(1+\delta\epsilon i)|X_j|^2\right\}X_j \nonumber  \\
   & + & e^{-i\alpha}\left[\frac{\mu}{N}\sum_{k=1}^N X_k+\frac{\nu}{N}\sum_{k=1}^{N}X_{N+k}\right] \label{eq:dXdt1} 
\end{eqnarray}
for $j=1,\ldots N$ and
\begin{eqnarray}
   \frac{dX_{j}}{dt} & = & i\omega X_{j}+\epsilon^{-1}\left\{1-(1+\delta\epsilon i)|X_j|^2\right\}X_j \nonumber \\
  & + & e^{-i\alpha}\left[\frac{\mu}{N}\sum_{k=1}^N X_{N+k}+\frac{\nu}{N}\sum_{k=1}^N X_k\right] \label{eq:dYdt1}
\end{eqnarray}
for $j=N+1,\ldots 2N$, where $X_j\in\mathbb{C}$, 
 and $\omega,\epsilon,\alpha,\mu$ and $\nu$ are real parameters.

These equations describe a pair of populations of $N$ Stuart-Landau 
oscillators
with all-to-all coupling within each population of strength
$\mu$, and all-to-all coupling between the two populations of strength $\nu$. Such oscillators are
related to the normal form of a Hopf bifurcation, and are a specific example of $\lambda-\omega$ 
oscillators~\cite{kuramotobook84,pikovskybook03,gre80}. Such a pair of coupled populations of oscillators
has been studied by several authors~\cite{abrmir08,barhun08,monkur04}, and can be thought of as the simplest
``network of networks'' that one could study.

Defining $X_j=r_je^{i\theta_j}$,
Eq.~(\ref{eq:dXdt1}) can be written
\begin{eqnarray}
    \frac{dr_j}{dt} & = & \epsilon^{-1}(1-r_j^2)r_j+\frac{\mu}{N}\sum_{k=1}^Nr_k\cos{(\theta_k-\theta_j-\alpha)} \nonumber \\
   & + & \frac{\nu}{N}\sum_{k=1}^{N}r_{N+k}\cos{(\theta_{N+k}-\theta_j-\alpha)} \label{eq:drdt} \\
    \frac{d\theta_j}{dt} & = & \omega-\delta r_j^2+\frac{1}{r_j}\left[\frac{\mu}{N}\sum_{k=1}^Nr_k\sin{(\theta_k-\theta_j-\alpha)} \right. \nonumber \\
   & + &\left.\frac{\nu}{N}\sum_{k=1}^{N}r_{N+k}\sin{(\theta_{N+k}-\theta_j-\alpha)}\right] \label{eq:dthetadt}
\end{eqnarray}
and Eq.~(\ref{eq:dYdt1}) can be written as a similar pair of equations. From Eq.~(\ref{eq:drdt}) we see that
as $\epsilon\rightarrow 0$, the rate of attraction to the limit cycle $r_j=1\ \forall j$
becomes infinite, and Eq.~(\ref{eq:dthetadt}) reduces to Equation~(1)
of~\cite{abrmir08} (after a redefinition of $\omega$), i.e.~our system reduces to a 
previously-studied network of phase oscillators. 
%Also, uncoupled oscillators (when $\mu=\nu=0$) have an attracting limit cycle of $r=1$ on which 
%$d\theta/dt=\omega$. The rate of attraction to this limit cycle is proportional to $\epsilon^{-1}$.
We will investigate the dynamics of~(\ref{eq:dXdt1})-(\ref{eq:dYdt1}) when $\epsilon\neq 0$. By allowing the
radius $r$ to vary, we expect a wider variety of behaviour than that seen in networks of phase oscillators;
for example, oscillator death and chaos~\cite{matmir91}.
For comparison
with previous results
we define $\beta=\pi/2-\alpha$ and we let $\mu=(1+A)/2, \nu=(1-A)/2$, where $A$ is a parameter~\cite{abrmir08}.

\begin{figure}
\includegraphics[width=3.4in]{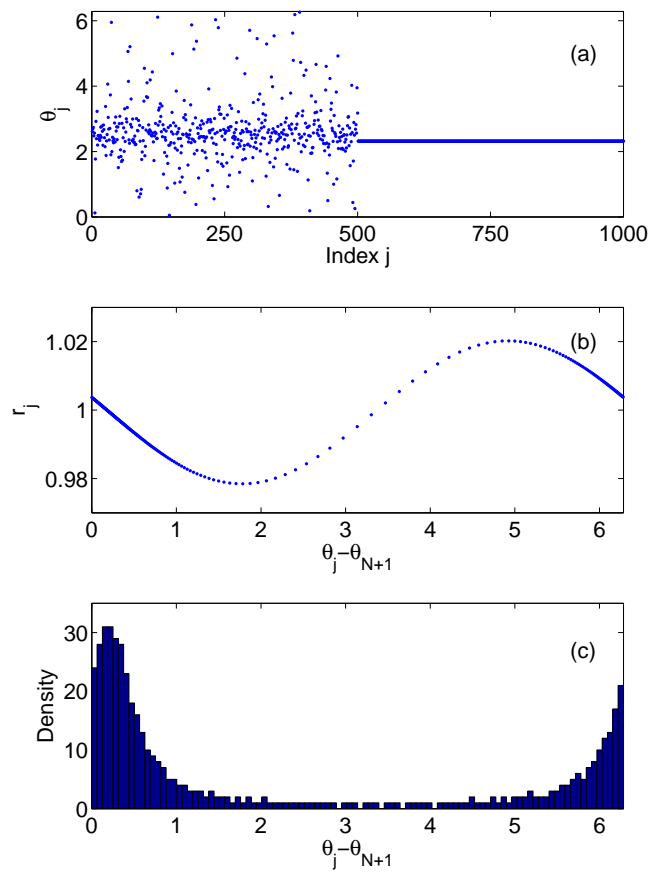}
\caption{(color online) A chimera state for~(\ref{eq:dXdt1})-(\ref{eq:dYdt1}). (a): A snapshot of the $\theta_j$. 
(b): $r_j$ as a function of $\theta_j$
(relative to $\theta_{N+1}$) for $j=1,\ldots N$. (c): The density of the $\theta_j$'s, relative to $\theta_{N+1}$,
for $j=1,\ldots N$. Parameters: $N=500, \omega=0,\epsilon=0.05, \beta=0.08,A=0.2,\delta=-0.1$.}
\label{fig:snap}
\end{figure}

Firstly, we show a chimera state for~(\ref{eq:dXdt1})-(\ref{eq:dYdt1});
see Fig.~\ref{fig:snap}. Panel (a) shows a snapshot of all $\theta_j$ at an arbitrary time. We see that 
population two (with $N+1\leq j\leq 2N$) has completely synchronised (all $r_j\approx 1.0019$), 
while oscillators in population one (with $1\leq j\leq N$) remain incoherent. Panel~(b) shows that oscillators in
population one lie 
on a closed curve (a slight distortion of the unit circle) in the complex plane. Panel (c) shows the angular
density of the oscillators in population one. It is non-uniform, i.e.~these oscillators are not completely
incoherent, and it was the dynamics of this density that~\citet{abrmir08} studied, using the parametrisation
of~\citet{ottant08}. 
In this chimera state the oscillators in population two have a constant angular velocity
and radius, and the distributions in panels (b) and (c) of Fig.~\ref{fig:snap} remain stationary.
It is worth noting that the chimera state shown in Fig.~\ref{fig:snap} is attracting, i.e.~nearby
states are attracted to it, unlike the corresponding chimera states in networks of identical
phase oscillators which are neutrally stable~\cite{abrmir08,pikros08}. Allowing both the radius and the phase
of the oscillators to vary seems to eliminate the non-generic behaviour seen in networks of identical,
sinusoidally-coupled phase oscillators, in the same way that making the oscillators non-identical 
does~\cite{lai09A,lai09B}.

We briefly digress to analyse the chimera state shown in Fig.~\ref{fig:snap} in the limit $\epsilon\rightarrow 0$, i.e.~$r_j=1 \forall j$. Let 
$\theta_j=\Theta$ for $N+1\leq j\leq 2N$ and move to a coordinate frame rotating with angular
velocity $\Omega$ in which $\Theta$ is
constant. Using rotational invariance, set $\Theta=0$. Then, from the equation for population two,
\begin{equation}
   0=\omega-\Omega-\delta-\mu\sin{\alpha}+\nu S \label{eq:phase}
\end{equation}
and (using Eq.~(\ref{eq:phase}))
each oscillator in population 1 satisfies
\begin{equation}
 %  \frac{d\theta}{dt}=\omega-\Omega-\delta+\sin{\alpha}-\nu\sin{(\theta+\alpha)}+\mu S\cos{\theta}-\mu C\sin{\theta)}
    \frac{d\theta}{dt}=\mu\sin{\alpha}-\nu S-\nu\sin{(\theta+\alpha)}+\mu S\cos{\theta}-\mu C\sin{\theta} \label{eq:dtheta}
\end{equation}
where $S\equiv N^{-1}\sum_{k=1}^N\sin{(\theta_k-\alpha)}$ and $C\equiv N^{-1}\sum_{k=1}^N\cos{(\theta_k-\alpha)}$.
In the limit $N\rightarrow\infty$, $S$ and $C$ are constant and can be replaced by the expected values of
$\sin{(\theta-\alpha)}$ and $\cos{(\theta-\alpha)}$ respectively, calculated using the angular density, 
$\rho(\theta)$, which is 
proportional to the
reciprocal of the velocity, $d\theta/dt$~\cite{abrstr06,shikur04}. Thus chimera states are described by the 
simultaneous solution of
\begin{equation}
   S = \int_0^{2\pi}\sin{(\theta-\alpha)}\rho(\theta)d\theta \label{eq:S}
\end{equation}
and
\begin{equation}
   C = \int_0^{2\pi}\cos{(\theta-\alpha)}\rho(\theta)d\theta \label{eq:C}
\end{equation}
where $\rho(\theta)=K(d\theta/dt)^{-1}$ and $K$ is a normalization 
factor such that $\int_0^{2\pi}\rho(\theta)d\theta=1$. Following solutions of Eqs.~(\ref{eq:S})-(\ref{eq:C})
as parameters are varied one can find regions of parameter space in which chimera states exist, 
in agreement with the results of~\citet{abrmir08} (results not shown).
Eq.~(\ref{eq:dtheta}) can be interpreted as 
saying that in a chimera state,
each oscillator in population one follows a periodic orbit, and is nonlinearly driven by its own mean field.
This effect is known to be capable of destroying completely synchronous behaviour~\cite{rospik07}.
We now analyse the chimera state in~(\ref{eq:dXdt1})-(\ref{eq:dYdt1}) for $\epsilon\neq 0$ using a similar argument, 
showing that it can be described by a single complex number.

Let $X_j=Y$ for $N+1\leq j\leq 2N$ and go to a rotating coordinate frame such
that $Y$ is constant in this frame. Rotate the frame so that $Y$ is real and positive. Then from Eq.~(\ref{eq:dYdt1}) we have
\begin{eqnarray}
   0 & = & i(\omega-\Omega) Y+\epsilon^{-1}\left\{1-(1+\delta\epsilon i)Y^2\right\}Y \nonumber \\
   & + & e^{-i\alpha}\left(\mu Y+\nu\widehat{X}\right) \label{eq:Y}
\end{eqnarray}
where $\widehat{X}\equiv N^{-1}\sum_{k=1}^N X_k$, 
and each oscillator in population one satisfies
\begin{eqnarray}
\frac{dX}{dt} & = & i(\omega-\Omega) X+\epsilon^{-1}\left\{1-(1+\delta\epsilon i)|X|^2\right\}X \nonumber  \\
   & + & e^{-i\alpha}\left[\mu\widehat{X}+\nu Y\right] \label{eq:dXdt2}
\end{eqnarray}
Given $\widehat{X}$, the real part of Eq.~(\ref{eq:Y}) can be solved for $Y$, and the imaginary 
part of Eq.~(\ref{eq:Y}) can be used 
to show that each oscillator in population one satisfies
\begin{eqnarray}
\frac{dX}{dt} & = & i\left[\delta Y^2+\mu\sin{\alpha}-(\nu/Y) \mbox{Im}\left\{e^{-i\alpha}\widehat{X}\right\}\right] X \nonumber \\
   & + & \!\!\epsilon^{-1}\left\{1\!-\!(1\!+\!\delta\epsilon i)|X|^2\right\}X\!+\!e^{-i\alpha}[\mu\widehat{X}\!+\!\nu Y] \label{eq:dXdt3}
\end{eqnarray}
i.e.~each oscillator in population one is driven in a nonlinear way by the mean field of population one.
Thus our self-consistency equation, i.e.~the analogue of~(\ref{eq:S})-(\ref{eq:C}), is
\begin{equation}
   \widehat{X}=\frac{1}{T(\widehat{X})}\int_0^{T(\widehat{X})}X(t;\widehat{X})\ dt \label{eq:Xhat}
\end{equation}
where $X(t;\widehat{X})$ is a periodic solution of Eq.~(\ref{eq:dXdt3}) with period $T(\widehat{X})$. 
The main difference between Eqs.~(\ref{eq:S})-(\ref{eq:C}) and Eq.~(\ref{eq:Xhat}) is that $X(t;\widehat{X})$ 
must be found by numerically integrating Eq.~(\ref{eq:dXdt3}) to find a periodic solution, whereas the 
periodic solution of Eq.~(\ref{eq:dtheta}) need not be found --- only the density, $\rho(\theta)$, proportional
to the reciprocal of the angular velocity, is needed.

\begin{figure}
\includegraphics[width=3.4in]{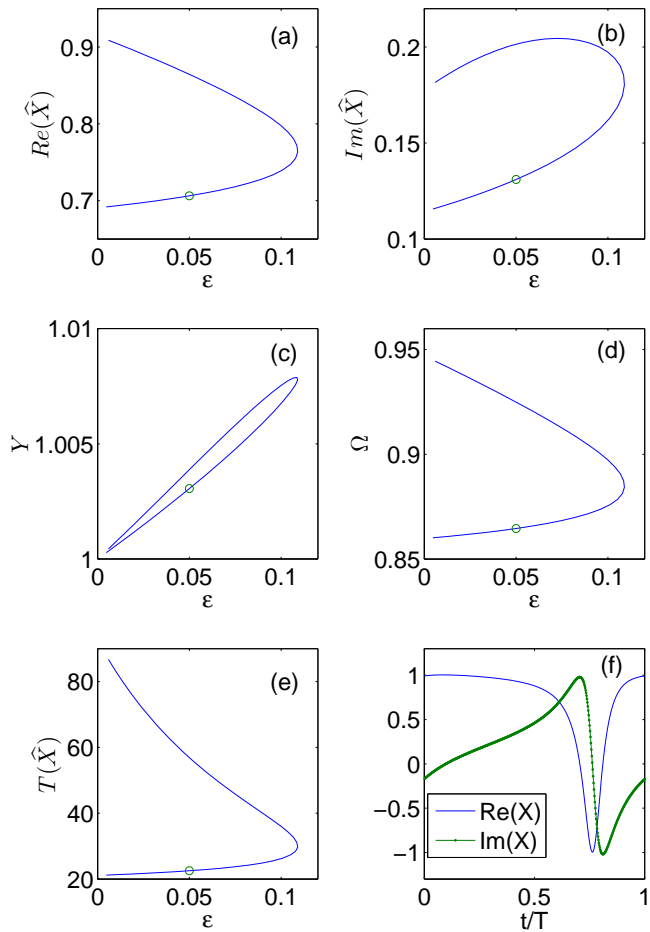}
\caption{(color online) The solution of Eq.~(\ref{eq:Xhat}). (a): Re($\widehat{X}$); (b): Im($\widehat{X}$);
(c): $Y$; (d): $\Omega$ and (e): $T(\widehat{X})$, as functions of $\epsilon$.
(f): Real and imaginary parts of the self-consistent solution
of Eq.~(\ref{eq:dXdt3}) for parameter values shown with a circle in panels (a)-(e).
Parameters: $\beta=0.08,A=0.2,\delta=-0.01$.}
\label{fig:Veps}
\end{figure}

Having
found a solution of Eq.~(\ref{eq:Xhat}), it can 
be numerically continued as parameters are varied.
Typical results are shown in Fig.~\ref{fig:Veps} where we vary $\epsilon$. We see that for these parameter
values the solution
of Eq.~(\ref{eq:Xhat}) can be continued to $\epsilon\approx 0.109$, where it appears to undergo a saddle-node
bifurcation. For $\epsilon$ small, points on the lower branch in panels (a)-(e) correspond to the stable
chimera known to exist~\cite{abrmir08} when $\epsilon=0$, while the upper branch corresponds to the 
saddle chimera.
A typical solution of Eq.~(\ref{eq:dXdt3}) is shown in Fig.~\ref{fig:Veps}~(f).

\begin{figure}
\includegraphics[width=3.4in]{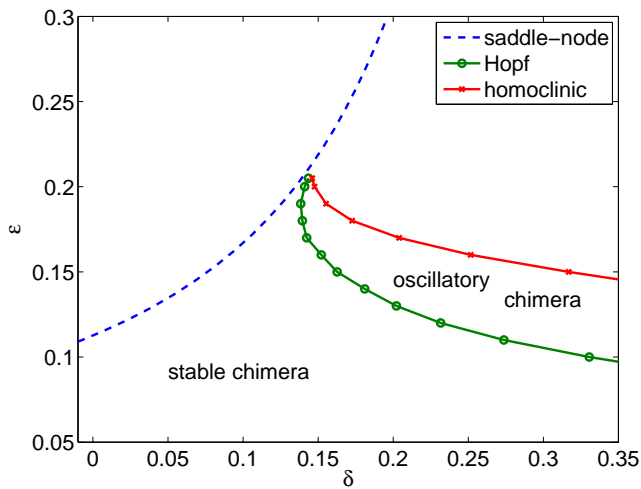}
\caption{(color online) Bifurcation curves in the $\delta-\epsilon$ plane for chimera solutions
of~(\ref{eq:dXdt1})-(\ref{eq:dYdt1}).
Hopf and homoclinic bifurcations were found by direct simulation of~(\ref{eq:dXdt1})-(\ref{eq:dYdt1}).
$A=0.2,\beta=0.08, N=500$.}
\label{fig:twopar}
\end{figure}

The saddle-node bifurcation seen in Fig.~\ref{fig:Veps} can be followed as a second parameter, say $\delta$, is
varied. The result is shown in Fig.~\ref{fig:twopar} (dashed curve). 
We see that as $\delta$ is increased, the range of values of $\epsilon$
for which a chimera state exists also increases. However, the curve of saddle-node bifurcations in 
Fig.~\ref{fig:twopar}
relates only to the {\em existence} of chimeras (found through a self-consistency argument 
similar to that of Kuramoto~\cite{kuramotobook84}) not their
stability. Numerical simulations of Eqs.~(\ref{eq:dXdt1})-(\ref{eq:dYdt1}) show that a stable stationary chimera
which exists to the right of the dashed curve in Fig.~\ref{fig:twopar}
can undergo a supercritical Hopf bifurcation as parameters are varied, leading to a ``breathing'' 
chimera~\cite{abrmir08,lai09A,lai09B}. These oscillatory states then seem to be destroyed in a homoclinic
bifurcation as parameters are further varied. Numerically determined curves of Hopf and homoclinic
bifurcations are shown in Fig.~\ref{fig:twopar}. These curves are conjectured to terminate at a Takens-Bogdanov bifurcation
on the curve of saddle-node bifurcations, which seems to be the generic arrangement
for chimera states~\cite{lai09B,abrmir08,mar10}.  Varying $A$ or $\beta$ rather than $\delta$ results in a similar
arrangement of saddle-node, Hopf and homoclinic bifurcation curves (results not shown).

To the left of the dashed curve in Fig.~\ref{fig:twopar} and above the curve of homoclinic bifurcations, 
the perfectly synchronous state ($X_j=X_k\ 
\forall j,k$) is stable. Despite the radii of our oscillators being able to vary, 
we have not been able to find oscillator death or more exotic dynamics by varying parameters.
Perhaps this is not too surprising, since non-identical oscillators (which we have not considered here)
and strong coupling relative to the attraction to the limit cycle (i.e.~the opposite limit from that considered
here) seem to be required to observe oscillator death~\cite{matmir91,erm90}.

In principle, the stability of the chimera states studied here, and thus the location of the Hopf 
bifurcation seen in Fig.~\ref{fig:twopar}, 
could be determined using the ideas
presented in Sec.~6 of~\citet{matmir91}. However, a difficulty arises because we do not have an analytic
expression for the chimera state around which to linearise --- the density, $\rho(r,\theta)$, 
of oscillators in population one
can only be found indirectly by numerically solving Eq.~(\ref{eq:dXdt3}). 
%Nevertheless, we have shown
%numerically that chimeras persist when the rate of attraction to the limit cycle $r=1$ is relaxed,
%in the same way that they persist for small heterogeneity in their intrinsic frequencies~\cite{lai09A,lai09B}.
(Note that the stability or otherwise of the periodic solution of Eq.~(\ref{eq:dXdt3}) that we find 
is not related to the
stability of the chimera state. Solving Eq.~(\ref{eq:dXdt3}) is just a convenient way of finding the invariant density
for population one.)

For chimeras to be observable in a physical system they must be generic, and not only occur in networks
of identical phase oscillators with all-to-all coupling via a sinusoidal function of phase differences, which
are known to have unusual properties~\cite{pikros08,marmir09,watstr93,watstr94}. Their persistence when
phase oscillators are made non-identical has been charcterised previously~\cite{lai09A,lai09B}, and in this Letter we have shown
that chimeras also persist (within limits) when both the amplitude and phase of the oscillators are allowed to vary.

One caveat is that the system studied here has all-to-all coupling, both within and between populations.
It would be interesting to determine whether this is necessary in order to observe chimeras. Indeed, this raises a more
general question as to which network topologies support chimeras. Also, the 
system~(\ref{eq:dXdt1})-(\ref{eq:dYdt1}) is invariant under the 
phase shift $X_j\mapsto e^{i\gamma}X_j \ \forall j$, where $\gamma$ is a real constant.
This seems
to be the reason that, in a chimera state, the synchronous population undergoes uniform rotation at fixed radius
in the complex plane, and we can describe the incoherent population as having a stationary distribution
in a uniformly rotating coordinate frame. It
would be of interest to study chimeras in networks for which this is not the case. Addressing these 
two issues would help determine the general robustness of chimeras, and thus the likelihood of them
having relevance to the physical world.

I thank Steve Strogatz for correspondence which inspired the work presented here.

%\bibliography{spiral}% Produces the bibliography via BibTeX.

\end{document}